# MAGNETIC DESIGN CONSTRAINTS OF HELICAL SOLENOIDS*

M. L. Lopes#, S. T. Krave, J. C. Tompkins, K. Yonehara, Fermi National Accelerator Laboratory, Batavia, IL 60510, USA
G. Flanagan, S. A. Kahn, Muons Inc., Batavia, IL, 60510, USA
K. Melconian, Texas A&M University, College Station, TX 77845 USA

*Abstract*

Helical solenoids have been proposed as an option for a Helical Cooling Channel for muons in a proposed Muon Collider. Helical solenoids can provide the required three main field components: solenoidal, helical dipole, and a helical gradient. In general terms, the last two are a function of many geometric parameters: coil aperture, coil radial and longitudinal dimensions, helix period and orbit radius. In this paper, we present design studies of a Helical Solenoid, addressing the geometric tunability limits and auxiliary correction system.

## INTRODUCTION

Helical cooling channels (HCC) based on a magnet system with a pressurized gas absorber in the aperture have been proposed as a highly efficient way to achieve 6D muon beam cooling [1-2]. The magnet system superimposes solenoid ($Bs$), helical dipole ($Bt$), and helical gradient ($G$) fields. The cooling channel was divided into several sections to provide the total phase space reduction of muon beams on the level of $10^5$-$10^6$, and to reduce the equilibrium emittance each consequent section has a smaller aperture and stronger magnetic fields. The field components ($Bt$ and $G$) are a function of many geometric parameters as it was presented in [3-5].

## GEOMETRICAL CONTRAINTS

A helical solenoid is defined essentially by four parameters: period ($\lambda$), orbit radius ($a$), aperture radius ($IR$) and the coil radial thickness ($DR$). Given a period $\lambda$, the orbit radius in a HCC is defined as:

$$a = \frac{\kappa \lambda}{2\pi} \qquad (1)$$

where $\kappa$ is a parameter for the cooling [1] and is normally equal to one [2].

The helical dipole and helical gradient are strongly dependent on the helical solenoid geometry. We can systematically study the effects of each one of these geometric parameters on the field components. For that, simulations using SolCalc [6] were carried out. The simulations assumed $\kappa$=1 and the solenoidal field component was normalized to 1. The aperture radius was varied from $IR$=0.1 to 0.4 m and the radial thickness was varied from $DR$=0.01 to 0.25 m for four values of $\lambda$: 0.4, 0.6, 0.8 and 1 m.

___________________
*Work supported in part by Fermi Research Alliance under DOE Contract DE-AC02-07CH11359.
#mllopes@fnal.gov

### Helical Dipole

The results of the helical dipole as function of $IR$ and $DR$ for different $\lambda$ show a strong dependence on geometry (figure 1). If we normalize the $IR$ and $DR$ by $a$, the results become independent of the period (figure 2), and they collapse to a single surface. A fit made to the surface of $Bt$ as function of $IR/a$ and $DR/a$ can be seen in figure 3.

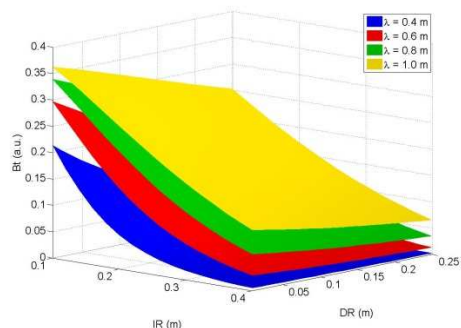

Figure 1: $Bt$ as function of $IR$, $DR$ and $\lambda$.

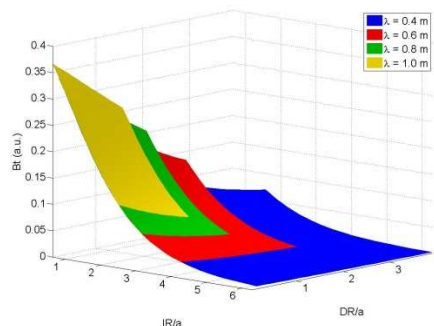

Figure 2: $Bt$ as function of $IR/a$, $DR/a$ and $\lambda$.

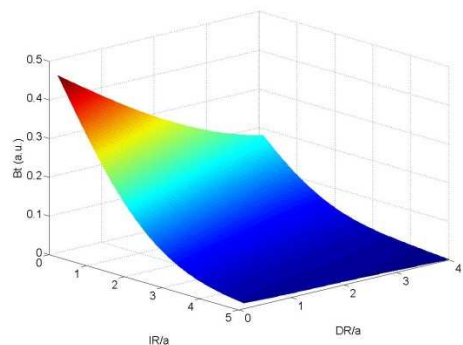

Figure 3: Fitted $Bt$ as function of $IR/a$ and $DR/a$

### Helical Gradient

The results of the helical gradient as a function of $IR$ and $DR$ for different $\lambda$ are shown in figure 4. As in the previous case, the dependence of *G* with the geometry is very strong. However, in order to make the results independent of the period, it is not enough to normalize *IR* and *DR* by *a* like in the case of the helical dipole. Since

$$G = \frac{dBt}{dr}, \quad (2)$$

$dr$ must also be normalized by *a*. Figure 5 shows the results of this normalization. Figure 6 shown the fitted surface of the normalized gradient (*G.a*) as function of *IR*/*a* and *DR*/*a*.

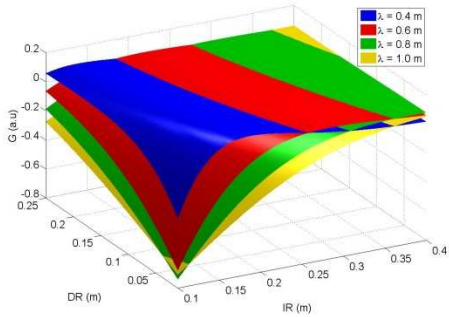

Figure 4: *G* as function of *IR*, *DR* and $\lambda$.

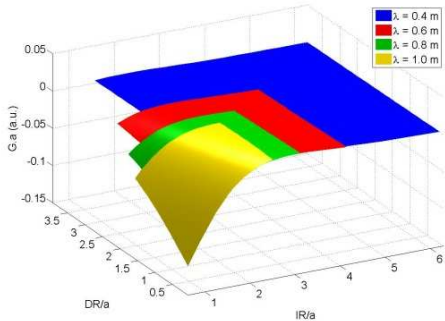

Figure 5: *G.a* as function of *IR*/*a*, *DR*/*a* and $\lambda$.

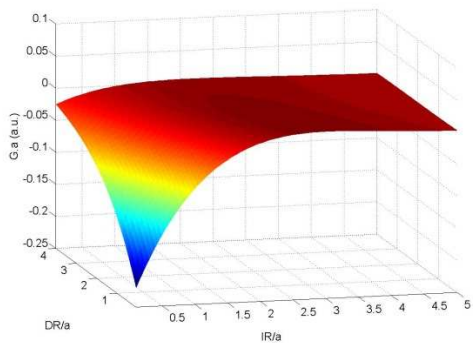

Figure 6: *G.a* as function of *IR*/*a* and *DR*/*a*.

## Peak Field

Another important parameter that limits the performance of a helical solenoid is the peak field on the coil. The same normalization discussed in the previous sections has been applied in order to have the results independent of $\lambda$. These results are shown in figure 7.

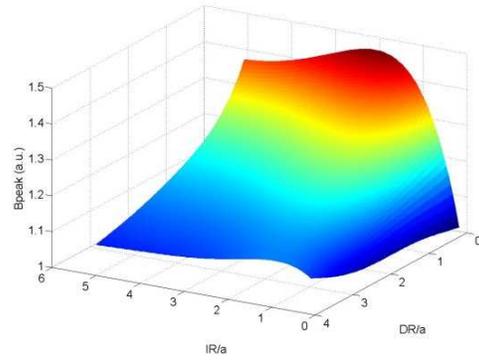

Figure 7: Normalized peak field as function of *IR*/*a* and *DR*/*a*.

## Current Density

Figure 8 shows the normalized (fitted) current density as function of *IR*/*a* and *DR*/*a*. The plot shows an exponential dependence of the current density with the coil thickness. This strong dependence may be the limiting factor for the overall performance of the coil.

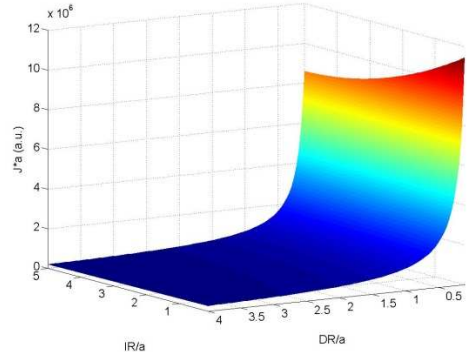

Figure 8: Normalized current density as function of *IR*/*a* and *DR*/*a*.

## FITTING

A 5$^{th}$ degree polynomial was used for the fitting presented in figure 3, 6 and 7. The fit is described by:

$$F(x,y) = \sum_{N=0}^{5} \sum_{M=0}^{5} pNM \cdot x^N y^M \quad (3)$$

with $N+M \leq 5$; where $x = IR/a$ and $y = DR/a$. The fitted coefficients for each case are presented in Table 1.

## THE DIPOLE AND GRADIENT RATIO

Achieving the desired dipole to gradient ratio is more important than obtaining the correct dipole or gradient individually. Once the ratio had been found, the current can be adjusted to get the correct dipole and gradient simultaneously. Using the fitted curves above, the limited range for *IR* and *DR* can quickly be determined for a desired *Bt*/*G.a* ratio (Figure 9). As an example, the range of coil thickness and inner radius for the high-field section of the helical solenoid, with a period of 0.4m (a=0.0637m) and a desired *Bt*/*G.a* of -14.2, is plotted in

Figure 10. This range may be further reduced by the limits of the superconductor [4].

Once the desired dipole and gradient components have been achieved, the resulting solenoid component will be larger than needed. The excess field can be reduced with an opposing external solenoid winding [3-4].

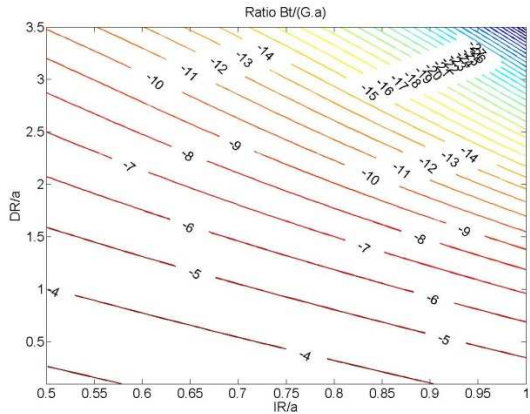

Figure 9: Contour plot of the ratio of the helical dipole and normalized helical gradient as a function of *IR/a* and *DR/a*

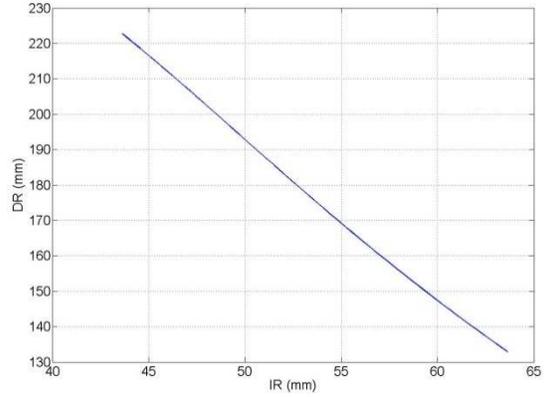

Figure 10: Coil thickness as function of inner radius, $\lambda$=400 mm, $B_z$=-17.3 T, $B_t$=4.06 T and $G$=-4.5 T/m.

Table 1: Coefficients for surface fit

| Coeff. | Dipole | Gradient | Peak field |
|---|---|---|---|
| **p00** | 4.90E-01 | -2.57E-01 | 9.46E-01 |
| **p10** | -1.76E-01 | 2.65E-01 | 7.44E-01 |
| **p01** | -9.67E-02 | 1.33E-01 | 4.37E-02 |
| **p20** | -1.11E-02 | -9.69E-02 | -3.62E-01 |
| **p11** | 7.31E-03 | -1.02E-01 | -3.01E-01 |
| **p02** | -1.28E-03 | -2.94E-02 | 2.23E-02 |
| **p30** | 1.40E-02 | 1.61E-02 | 7.88E-02 |
| **p21** | 1.03E-02 | 2.79E-02 | 9.56E-02 |
| **p12** | 8.35E-03 | 1.62E-02 | 6.84E-02 |
| **p03** | 2.55E-03 | 2.78E-03 | -2.88E-02 |
| **p40** | -2.30E-03 | -1.19E-03 | -8.14E-03 |
| **p31** | -2.40E-03 | -3.29E-03 | -1.31E-02 |
| **p22** | -2.19E-03 | -2.79E-03 | -1.02E-02 |
| **p13** | -1.50E-03 | -1.15E-03 | -1.06E-02 |
| **p04** | -3.24E-04 | 6.83E-06 | 9.39E-03 |
| **p50** | 1.22E-04 | 2.87E-05 | 3.29E-04 |
| **p41** | 1.52E-04 | 1.42E-04 | 6.42E-04 |
| **p32** | 1.45E-04 | 1.56E-04 | 5.62E-04 |
| **p23** | 1.69E-04 | 1.00E-04 | 6.38E-04 |
| **p14** | 4.27E-05 | 2.07E-05 | 5.98E-04 |
| **p05** | 2.99E-05 | -1.22E-05 | -9.23E-04 |

## FUTURE STUDIES

Alternative correction and winding schemes are being investigated to eliminate or reduce the required correction solenoid field. Examples include tilting the main coils or adding a few individual helical windings to increase the dipole and gradient component.

## CONCLUSION

The paper presented the relationship between the field components as a function of geometric parameters. The helical dipole and helical gradient show a strong dependence on the coil radial thickness and aperture radius. A normalized surface was found for the dipole, gradient, and peak field independent of $\lambda$. From these results, a range for the aperture size and coil thickness can be found without running magnetic simulations. Further limits from the coil peak field and current density still need to be considered.